\documentstyle[prl,aps,multicol,epsfig]{revtex}
\RequirePackage{rawfonts}

\begin{document}

\title{Multi-band Gutzwiller wave functions for itinerant 
ferromagnetism\footnote{Dedicated to Martin C.~Gutzwiller on 
the occasion of his 75th birthday.}}

\author{J\"org~B\"unemann and Florian~Gebhard}
\address{Fachbereich Physik, Philipps-Universit\"at Marburg, D-35032 
Marburg, Germany}
\author{Werner Weber}
\address{Institut~f\"ur~Physik, Universit\"at Dortmund, 
D-44221 Dortmund, Germany}

\maketitle

\begin{abstract}
Multi-band Gutzwiller-correlated wave functions 
reconcile the contrasting concepts of itinerant band electrons
versus electrons localized in partially filled atomic shells. 
The approximate evaluation  
of these variational ground states becomes exact in the limit
of large coordination number. The result allows the 
identification of quasi-particle band structures
for correlated electron systems.
As a first application, we summarize a study of itinerant ferromagnetism in a two-band
model, thereby elucidating the co-operation of the Coulomb repulsion and the
Hund's-rule exchange. Then, we present results of calculations for 
ferromagnetic nickel,
using a realistic 18 spin-orbital basis of $4s$, $4p$ and $3d$ valence electrons.
Good agreement with the experimental ground-state properties of nickel is
obtained. In particular, the quasi-particle energy bands agree much better
with the photo-emission and Fermi surface data than the band structure
obtained from spin-density functional theory.
Finally, we present results for the variational spinwave dispersion
for our two-band model.

\end{abstract}


\begin{multicols}{2}
\narrowtext

\section{Introduction}

How well do we understand the itinerant ferromagnetism
of iron, nickel, and other transition metals and their compounds?

More than 50~years ago two basically different scenarios
had emerged from early quantum-mechanical considerations
on electrons in metals with partly filled $d$~bands.
In scenario~I, suggested by Slater~\cite{Slaterearly} and 
Stoner~\cite{Stoner}, 
band theory alone was proposed to account for itinerant
ferromagnetism:
due to the Pauli principle, electrons with parallel spins cannot come
arbitrarily close to each other (``Pauli'' or ``exchange hole''), 
and, thus, a ferromagnetic alignment of the electron spins
reduces the total Coulomb energy with respect to the
paramagnetic situation (``exchange energy'').
This scenario, equivalent to Hartree-Fock theory,
was criticized by van Vleck~\cite{vanVleck}.
In an independent-electron theory electrons are distributed statistically
over the lattice which implies strong charge fluctuations on
the transition metal atoms.
 
Consequently, as a scenario~II, 
van Vleck~\cite{vanVleck} emphasized
the importance of electron correlations:
due to the strong electron-electron interactions,
charge fluctuations in the atomic $d$~shells are strongly suppressed
(``minimum polarity model''),
and, thus, atomic rather than band aspects are decisive
for itinerant ferromagnetism. In scenario~II, 
the atomic magnetic moments arise due to
the local Coulomb interaction (in particular, 
Hund's-rule couplings), and they may
align because of the electrons' motion through the crystal.

In principle, such a dispute can be resolved in natural sciences.
The corresponding theories have to be worked out in detail, and their
results and predictions have to be compared to experiments.
This was indeed done for scenario~I~\cite{Moruzzi}.
The (spin-)density functional theory
is a refined band theory which describes
simple metals with considerable
success. Unfortunately, progress for scenario~II was much slower.
It calls for a theory of correlated electrons, i.e.,
a genuine many-body problem has to be solved.
It was only recently that reliable theoretical tools became available
which allow to elucidate scenario~II in more 
detail~\cite{Nolting,Hasegawa,Pruschke,Vollhardt,BGWvoll}.

A first step in this direction was the
formulation of appropriate model Hamiltonians which
allowed to discuss matters concisely. In 1963/1964
Gutzwiller~\cite{GutzPRL63,Gutzwiller1964}, 
Hubbard~\cite{Hubbard}, and Kanamori~\cite{Kanamori}
independently introduced the Hubbard model
which since then has become the standard model
for correlated electrons on a lattice.
This model covers both aspects of $d$~electrons on a lattice:
they can move through the crystal, and they strongly interact
when they sit on the same lattice site.
The model is discussed in more detail in Sec.~\ref{Hubbardmodel}.

Even nowadays, it is impossible to calculate exact ground-state properties
of such a model in three dimensions. 
In 1963/1964 Gutzwiller introduced a 
trial state to examine variationally the possibility of ferromagnetism
in such a model~\cite{GutzPRL63,Gutzwiller1964}. 
His wave function covers both limits of
weak and strong correlations and should, therefore, be suitable
to provide qualitative insights into the magnetic phase diagram
of the Hubbard model. 
Gutzwiller-correlated wave functions for multi-band Hubbard models
are defined and analyzed in Sec.~\ref{GWF}.

As envisaged by van Vleck~\cite{vanVleck},
Gutzwiller~\cite{GutzPRL63}
found that ferromagnetism is non-generic in the
one-band Hubbard model, in contrast to the predictions of the
Stoner theory.
The reason is simple: the Gutzwiller wave function not only describes
the exchange hole between electrons of the same spin species
but also the ``correlation hole'' between electrons of 
different spin species which forms when the electron-electron interaction
is taken into account consistently. Therefore, the energy difference
between the correlated paramagnetic and ferromagnetic states is
very small, and paramagnetism wins in almost all one-band cases.
In the multi-band situation~\cite{Gutzwiller1964}, 
itinerant ferromagnetism is favored by
two concomitant effects: 
(i)~the suppression of atomic charge fluctuations
such that, (ii), the local exchange interactions give rise to
a local magnetic moment as in isolated atoms (Hund's rule).
Hence, as proposed by van~Vleck, small charge fluctuations
{\sl and\/} local exchange interactions
generate large local moments which are present
in both the paramagnetic and ferromagnetic phases.
At low temperatures the electrons' motion through the crystal may then
lead to long-range order of these pre-formed local moments.
In this way, Gutzwiller's approach unified and substantiated
van~Vleck's early ideas
on a correlated-electron theory of itinerant ferromagnetism.

Unfortunately, the evaluation of multi-band Gutzwiller
wave functions itself poses a most difficult many-body problem.
Perturbative treatments~\cite{Stollhoff,Carmelo} are constrained
to small to moderate interaction strengths. The region of
strong correlations could only be addressed within the 
``Gutzwiller approximation''~\cite{GutzPRL63,Gutzwiller1964,Vollirev}
and its various extensions~\cite{Chao,Fazekas}.
Thus, an application to ferromagnetism of real materials 
was not in sight,
and Gutzwiller's qualitative findings were not appreciated
much in the sixties.

About 25~years later, a major technical advancement opened
a new way to analyze Gutzwiller-correlated wave functions.
The one-band Gutzwiller wave function was evaluated exactly 
in one dimension~\cite{MVGUTZ,GVGUTZ}, and
the Gutzwiller approximation was found 
to become exact for the one-band Gutzwiller wave function
in the limit of infinite spatial dimensions, 
$d\to\infty$~\cite{MVGUTZ,MVPRLdinfty,BUCH}.
Based on a diagrammatical approach, Gebhard~\cite{Geb1990} developed
a compact formalism which allows to calculate the variational
ground-state energy in infinite dimensions without
the evaluation of a single diagram. He found that $1/d$~corrections
are quantitatively small for the Gutzwiller wave function, 
i.e., the result in infinite dimensions
is a reliable approximation to three dimensions.

Recently, Gebhard's approach was generalized by us 
to the case of multi-band
Gutzwiller wave functions~\cite{BGWvoll}. Thereby, earlier results
by B\"unemann and Weber~\cite{BueWeb}, based on 
a generic extension of the Gutzwiller
approximation~\cite{JoergEJB}, were found to 
become exact in infinite dimensions~\cite{BGWhalb}.
Half a century after van Vleck's call for new many-particle
methods and more than thirty years after Gutzwiller's 
first paper on the subject, techniques are at hand to
carry out the program outlined above as scenario~II.

Our results on a two-band toy model in Sect.~\ref{toymodel}
confirm Gutzwiller's insights on the fundamental requirements
for itinerant ferromagnetism. Substantial atomic
Coulomb interactions suppress local charge fluctuations and thereby 
allow the generation of local moments due to 
the atomic Hund's-rule exchange couplings.
Our quantitative results from
a full-scale calculation for nickel in Sect.~\ref{nickel}
confirm these qualitative findings.
The quasi-particle properties of this material (e.g., exchange splittings,
density of states, topology of the Fermi surface) are reproduced
quantitatively from our Gutzwiller-correlated wave functions.
In particular, strong electron correlations are required to
describe nickel appropriately.

In Sect.~\ref{spinwave}, we briefly discuss
a new method to calculate spinwave spectra
of correlated electron systems based on
the Gutzwiller variational scheme~\cite{Buene2000}.
In this way, the dispersion relation of the
fundamental low-energy excitations can be derived
consistently. Albeit the description is based on
itinerant electrons, the results for strong ferromagnetism
in a generic two-band
model resemble those of a Heisenberg model for localized
spins. Again, the Gutzwiller approach provides a comprehensive
scheme for a unified description of localized and itinerant
aspects of transition metals.

In this brief review we skip all technical details
and refer the reader to the original literature
for further information.

\section{Hamilton operator}
\label{Hubbardmodel}


Our multi-band Hubbard model~\cite{Hubbard} is defined by the Hamiltonian 
\begin{equation}
\hat{H}=\sum_{i,j;\bbox{\sigma},\bbox{\sigma'}}
t_{i,j}^{\bbox{\sigma},\bbox{\sigma'}}
\hat{c}_{i;\bbox{\sigma}}^{+}
\hat{c}_{j;\bbox{\sigma'}}^{\vphantom{+}}
+ \sum_i\hat{H}_{i;\text{at}}\equiv \hat{H}_1+\hat{H}_{\text{at}}\;.  \label{1}
\end{equation}
Here, $\hat{c}_{i;\bbox{\sigma}}^{+}$ creates an electron with combined
spin-orbit index~$\bbox{\sigma}=1,\ldots ,2N$ ($N=5$ for 3$d$~electrons) at
the lattice site~$i$ of a solid. 

The most general case is treated in Ref.~\cite{BGWvoll}.
In this work we assume for simplicity that different types
of orbitals belong
to different representations of the point group of the respective 
atomic state (e.g., $s$, $p$, $d(e_g)$, $d(t_{2g})$).
In this case, different types of orbitals do not mix locally, and, thus,
the crystal field is of the from
$t_{i,i}^{\bbox{\sigma},\bbox{\sigma'}}= \epsilon_{\bbox{\sigma}}
\delta_{\bbox{\sigma},\bbox{\sigma'}}$. 
Consequently, we use 
normalized single-particle product states $|\Phi_0\rangle$ 
which respect the symmetry of the lattice, i.e.,
\begin{equation}
\langle \Phi_0 | \hat{c}_{i;\bbox{\sigma}}^{+}
\hat{c}_{i;\bbox{\sigma'}}^{\vphantom{+}} | \Phi_0 \rangle
= \delta_{\bbox{\sigma}, \bbox{\sigma'}} n_{i;\bbox{\sigma}}^0\; .
\end{equation}
We further assume that the local interaction
is site-independent
\begin{equation}
\hat{H}_{i;\text{at}} =
\sum_{\bbox{\sigma_1},\bbox{\sigma_2},\bbox{\sigma_3},\bbox{\sigma_4}}
{\cal U}^{\bbox{\sigma_1},\bbox{\sigma_2};\bbox{\sigma_3},\bbox{\sigma_4}}
\hat{c}_{i;\bbox{\sigma_1}}^{+}\hat{c}_{i;\bbox{\sigma_2}}^{+}
\hat{c}_{i;\bbox{\sigma_3}}^{\vphantom{+}}
\hat{c}_{i;\bbox{\sigma_4}}^{\vphantom{+}}\;.  \label{fullHat}
\end{equation}
This term represents all possible local Coulomb interactions.


As our basis for the atomic problem we choose the configuration states
\begin{eqnarray}
|I\rangle &=& |\bbox{\sigma_1},\bbox{\sigma_2}, \ldots\rangle \nonumber \\
&= &\hat{c}_{i;\bbox{\sigma_1}}^{+}\hat{c}_{i;\bbox{\sigma_2}}^{+} \cdots
|{\rm vaccum}\rangle \quad (\bbox{\sigma_1}<\bbox{\sigma_2}< \cdots)
\; ,
\end{eqnarray}
which are the ``Slater determinants'' in atomic physics.
The diagonalization of the Hamiltonian~$\hat{H}_{i;{\rm at}}$
is a standard exercise~\cite{Sugano}.
The eigenstates~$|\Gamma\rangle$ obey
\begin{equation}
|\Gamma \rangle =\sum_I T_{I,\Gamma }|I\rangle \;,
\end{equation}
where $T_{I,\Gamma }$ are the elements of the unitary matrix which
diagonalizes the atomic Hamiltonian matrix with entries
$\langle I | \hat{H}_{i;{\rm at}}| I'\rangle$.
Then, 
\begin{eqnarray}
\hat{H}_{i;{\rm at}} &=& \sum_{\Gamma} E_{\Gamma}\hat{m}_{\Gamma}\; ,\\
\hat{m}_{\Gamma} &=& |\Gamma\rangle\langle \Gamma | \; .
\end{eqnarray}
The atomic properties, i.e., eigenenergies $E_{\Gamma}$, 
eigenstates~$|\Gamma\rangle$, and
matrix elements~$T_{I,\Gamma}$, are essential ingredients of our 
solid-state theory.

\section{Multi-band Gutzwiller wave functions}
\label{GWF}


Gutzwiller-correlated wave functions 
are written as a many-particle correlator~$\hat{P}_{\text{G}}$ acting
on a normalized single-particle product state~$|\Phi_0\rangle $, 
\begin{equation}
|\Psi_{\text{G}}\rangle =\hat{P}_{\text{G}}|\Phi_0\rangle \;.
\end{equation}
The single-particle wave function $|\Phi_0\rangle$
contains many configurations which are
energetically unfavorable with respect to the atomic interactions. 
Hence, the correlator~$\hat{P}_{\text{G}}$ is chosen to 
suppress the weight of these configurations to minimize the total 
energy in~(\ref{1}). In the limit of strong correlations the Gutzwiller 
correlator~$\hat{P}_{\text{G}}$ 
should project onto atomic eigenstates. Therefore, the proper multi-band
Gutzwiller wave function with atomic correlations reads 
\begin{eqnarray}
\hat{P}_{\text{G}} &=& \prod_i\hat{P}_{i;\text{G}} \; , \nonumber \\
\hat{P}_{i;\text{G}} 
&=&
\prod_{\Gamma} \lambda_{i;\Gamma }^{\hat{m}_{i;\Gamma}}
=\prod_{\Gamma} \left[ 1+\left( \lambda_{i;\Gamma }-1\right) \hat{m}_{i;\Gamma}
\right]  \label{GutzcorrdegbandsHund} \\
&=& 1+\sum_{\Gamma} 
\left( \lambda_{i;\Gamma}-1\right) \hat{m}_{i;\Gamma }
\;. \nonumber 
\end{eqnarray}
The $2^{2N}$ variational parameters~$\lambda_{i;\Gamma }$ per site are
real, positive numbers. For $\lambda_{i;\Gamma_0}\neq 0$ and all other 
$\lambda_{i;\Gamma }= 0$ all atomic configurations at 
site~$i$ but~$|\Gamma_0\rangle $ are removed from~$|\Phi_0\rangle $. 
Therefore, by construction, $|\Psi_{\rm G}\rangle$ covers both limits
of weak and strong coupling. In this way it incorporates both itinerant
and local aspects of correlated electrons in narrow-band systems.


The class of Gutzwiller-correlated wave functions as specified 
in~(\ref{GutzcorrdegbandsHund}) was evaluated exactly in the limit of
infinite dimensions in Ref.~\cite{BGWvoll}. 
The expectation value of the Hamiltonian~(\ref{1}) reads
\begin{eqnarray}
\langle \hat{H} \rangle &=& 
\frac{\langle \Psi_{\rm G} | \hat{H} |\Psi_{\rm G} \rangle }%
{\langle \Psi_{\rm G} | \Psi_{\rm G} \rangle } \nonumber \\
&=&
\sum_{i\neq j;\bbox{\sigma},\bbox{\sigma'}}
t_{i,j}^{\,\bbox{\sigma},\bbox{\sigma'}} \sqrt{q_{i;\bbox{\sigma}}}
\sqrt{q_{j;\bbox{\sigma'}}}
\langle \Phi_0  |
\hat{c}_{i;\bbox{\sigma}}^{+}
\hat{c}_{j;\bbox{\sigma'}}^{\vphantom{+}}
| \Phi_0 \rangle \label{allresultsdegbandsHund}\\
&& +\sum_{i;\bbox{\sigma}}\epsilon_{\bbox{\sigma}}n_{i;\bbox{\sigma}}^0
+\sum_{i;\Gamma }E_{\Gamma}m_{i;\Gamma }\;. \nonumber 
\end{eqnarray}
Here, $n_{i,\bbox{\sigma}}^0=\langle \Phi_0 | \hat{n}_{i;\bbox{\sigma}}|
\Phi_0\rangle $ is the local particle density
in $| \Phi_0 \rangle$. 
The local $q$-factors are given by~\cite{BGWvoll}
\begin{eqnarray}
\sqrt{q_{\bbox{\sigma}}}  &=& 
\sqrt{\frac{1}{n_{\bbox{\sigma}}^{0}(1-n_{\bbox{\sigma}}^{0})}} 
\sum_{\Gamma,\Gamma'} 
\sqrt{\frac{m_{\Gamma}m_{\Gamma'}}{m_{\Gamma}^0m_{\Gamma'}^0}}  
\sum_{ I,I' \, (\bbox{\sigma}\not\in I,I')} 
f_{\bbox{\sigma}}^I f_{\bbox{\sigma}}^{I'} 
\nonumber \\
&&  \sqrt{ m_{(I'\cup\bbox{\sigma})}^{0} m_{I'}^{0} }
T_{\Gamma,(I\cup\bbox{\sigma})}^+ T_{(I'\cup\bbox{\sigma}),\Gamma}
T_{\Gamma',I'}^+ T_{I,\Gamma'} \; , \label{qfactor}
\end{eqnarray}
where $m_{i;I}^{0}$ ($m_{i;\Gamma}^{0}$)
is the probability to find the configuration~$|I\rangle$ (the atomic
eigenstate~$|\Gamma\rangle$)
on site~$i$ in the single-particle product state~$|\Phi_0\rangle$.
The fermionic sign function
\begin{equation}
f_{\bbox{\sigma}}^{I}  \equiv \langle I\cup \bbox{\sigma} 
| \hat{c}_{\bbox{\sigma}}^+  | I\rangle
\end{equation}
gives a minus (plus) sign if it takes an odd (even) number of anticommutations
to shift the operator $\hat{c}_{\bbox{\sigma}}^+$
to its proper place in the sequence of electron creation operators
in $|I\cup \bbox{\sigma} \rangle$.

Eqs.~(\ref{allresultsdegbandsHund}) and~(\ref{qfactor}) show that we may
replace the original variational parameters~$\lambda_{i;\Gamma }$ by their
physical counterparts, the atomic occupancies~$m_{i;\Gamma }$.
They are related by the simple equation~\cite{BGWvoll}
\begin{equation}
m_{i;\Gamma } = \lambda_{i;\Gamma }^2 m_{i;\Gamma }^0 \; .
\label{HundgutzrelationsA}
\end{equation}
The probability for an empty site ($|I|=0$) is obtained
from the completeness condition,
\begin{equation}
m_{i;\emptyset }=1-\sum_{\Gamma\,(|\Gamma |\geq 1)}m_{i;\Gamma }\;.  
\label{completenessgamma}
\end{equation}
The probabilities for a singly occupied site ($|I|=1$) are given by
\begin{mathletters}
\label{nGutz}
\begin{eqnarray}
m_{i;\bbox{\sigma}} &=& n_{i;\bbox{\sigma}}^0 -
\sum_{I \, (|I|\geq 2) \, (\bbox{\sigma}\in I) } m_{i;I} \;, \\
m_{i;I} &=& \sum_{K} \biggl|
\sum_{\Gamma} 
\sqrt{\frac{m_{i;\Gamma}}{m_{i;\Gamma}^0}}
T_{\Gamma,I}^+ T_{K,\Gamma}
\biggr|^2 m_{i;K}^0 \; .  \label{mImGammastrich}
\end{eqnarray}\end{mathletters}%
The parameters $m_{i;\emptyset}$ and $m_{i;\bbox{\sigma}}$
must not be varied independently.
All quantities in~(\ref{allresultsdegbandsHund}) are now expressed in
terms of the atomic multi-particle
occupancies~$m_{i;\Gamma}$ ($|\Gamma|\geq 2$),
the local densities $n_{i;\bbox{\sigma}}^0$, and further variational
parameters in~$|\Phi_0\rangle$.

It is seen that the variational ground-state energy can be cast into the
form of the expectation value of an effective single-particle Hamiltonian
with renormalized electron transfer amplitudes
$\widetilde{t}_{i,j}^{\,\bbox{\sigma},\bbox{\sigma'}}$,
\begin{eqnarray}
\hat{H}_{\rm eff} &=& 
\!\!\sum_{i\neq j;\bbox{\sigma},\bbox{\sigma'}}
\widetilde{t}_{i,j}^{\,\bbox{\sigma},\bbox{\sigma'}} 
\hat{c}_{i;\bbox{\sigma}}^{+}
\hat{c}_{j;\bbox{\sigma'}}^{\vphantom{+}}
+\sum_{i;\bbox{\sigma}}\epsilon_{\bbox{\sigma}}\hat{n}_{i;\bbox{\sigma}}
+\sum_{i;\Gamma }E_{\Gamma}m_{i;\Gamma }\, , \nonumber \\
\widetilde{t}_{i,j}^{\,\bbox{\sigma},\bbox{\sigma'}} 
&=& 
\sqrt{q_{i;\bbox{\sigma}}}
\sqrt{q_{j;\bbox{\sigma'}}}
t_{i,j}^{\,\bbox{\sigma},\bbox{\sigma'}}  \; .
\label{widetildet}
\label{Heff}
\end{eqnarray}
Therefore, $|\Phi_0\rangle$ is the ground state of $\hat{H}_{\rm eff}$
whose parameters have to be determined self-consistently
from the minimization of $\langle \Phi_0| \hat{H}_{\rm eff}| \Phi_0 \rangle$ 
with respect to $m_{i;\Gamma}$ and $n_{i;\bbox{\sigma}}^0$.
For the optimum set of parameters, $\hat{H}_{\rm eff}^{\rm opt}$ 
{\sl defines\/} a band structure for {\sl correlated\/} electrons. 
Similar to density-functional theory, this interpretation
of our ground-state results opens the way to detailed comparisons with 
experimental results.
 
Applications are discussed in the next two sections.

\section{Results for a generic two-band model}
\label{toymodel}

The atomic Hamiltonian for a two-band model ($b=1,2$)
can be cast into the form 
\begin{eqnarray}
\hat{H}_{i;\text{at}} &=&
U \sum_{b}\hat{n}_{b,\uparrow}\hat{n}_{b,\downarrow}
+U'\sum_{\sigma,\sigma'}\hat{n}_{1,\sigma}\hat{n}_{2,\sigma'}
-J\sum_{\sigma}\hat{n}_{1,\sigma}\hat{n}_{2,\sigma}
\nonumber 
\\[3pt]
&& +J\sum_{\sigma}\hat{c}_{1,\sigma}^{+}
\hat{c}_{2,-\sigma}^{+}
\hat{c}_{1,-\sigma}^{\vphantom{+}}
\hat{c}_{2,\sigma}^{\vphantom{+}}  \label{twoorbhamiltonian} \\
&& +J_{\text{C}} \Bigl(
\hat{c}_{1,\uparrow}^{+}\hat{c}_{1,\downarrow}^{+}
\hat{c}_{2,\downarrow}^{\vphantom{+}}\hat{c}_{2,\uparrow}^{\vphantom{+}}
+ 
\hat{c}_{2,\uparrow}^{+}\hat{c}_{2,\downarrow}^{+}
\hat{c}_{1,\downarrow}^{\vphantom{+}}\hat{c}_{1,\uparrow}^{\vphantom{+}}
\Bigr)\;.  \nonumber
\end{eqnarray}
For two orbitals, $\hat{H}_{\text{at}}$ exhausts all possible
two-body interaction terms.

We assume that the model describes two degenerate $d(e_g)$ orbitals
which leads to the following restrictions enforced by the cubic 
symmetry~\cite{Sugano}:
(i)~$J=J_{\text{C}}$, and (ii)~$U-U'=2J$.
Therefore, there are two independent Coulomb parameters,
the local Coulomb repulsion $U$ and the local exchange coupling~$J$.
Since $U, U'={\cal O}(10\, {\rm eV})$
are of the same order of magnitude the relation~(ii) shows
that $J={\cal O}(1\, {\rm eV})$ is of the typical strength
of the atomic Hund's rule couplings.

For the one-particle part~$\hat{H}_1$ we 
use an orthogonal tight-binding Hamiltonian with first and second
nearest neighbor hopping matrix elements. 
We apply the two-center approximation for the 
hopping matrix elements and exclude any spin-flip
hopping.
The resulting band structure (width~$W=6.6\, {\rm eV}$) is discussed
in detail in Ref.~\cite{BGWvoll}.

In the following we concentrate on two band-fillings~$0\leq n\leq 4$, 
the number of electrons per lattice site. Alternatively,
we use $0\leq n_{\bbox{\sigma}}=n/4 \leq 1$, 
the particle density per band and per spin direction.
For~$n_{\bbox{\sigma}}=0.29$, the non-interacting density 
of states has a maximum 
which is most favorable for ferromagnetism according to the 
Stoner (i.e, Hartree-Fock) theory.
Moreover, we study~$n_{\bbox{\sigma}}=0.35$, 
where the density-of-states has a positive curvature.

\begin{figure}[h]
\vspace*{-20mm}
  \begin{center}
    \epsfig{file=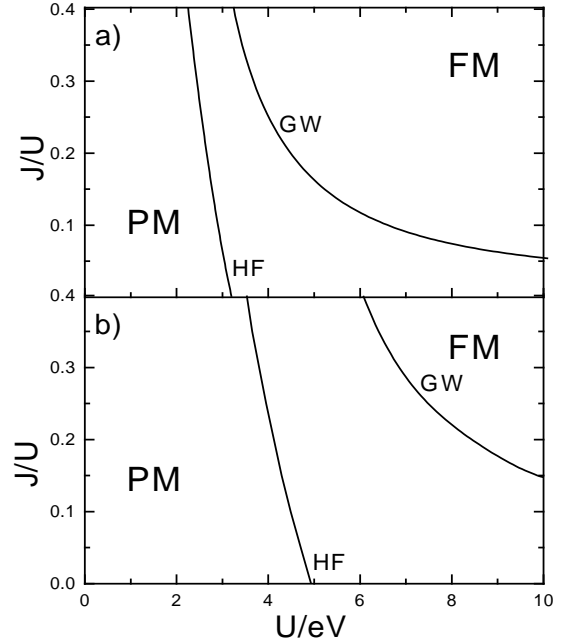,width=8cm,height=11cm}
  \end{center}
\vspace*{-11mm}
\caption{Phase diagram as a function of~$U$ and~$J$ for the
Hartree-Fock--Stoner theory (HF) and the Gutzwiller wave function
(GW) for (a)~$n/4=0.29$ and (b)~$n/4=0.35$; 
PM: paramagnet, FM: ferromagnet.}
\label{fig:phasediaferro}
\end{figure}

In Fig.~\ref{fig:phasediaferro} 
we display the $J$-$U$ phase diagram for both
fillings. It shows that Hartree--Fock theory always predicts a ferromagnetic
instability. In contrast, the correlated-electron approach strongly supports
the ideas of van Vleck~\cite{vanVleck} and Gutzwiller~\cite{Gutzwiller1964}:
(i)~a substantial on-site exchange~$J$ is required for the occurrence
of ferromagnetism if, (ii), realistic Coulomb repulsions~$U$
are assumed. In~1964 Gutzwiller wrote
``\ldots {\sl one may hope to show some
day that the terms~(ii) can never induce ferromagnetism at all by themselves}
\ldots''. Fig.~\ref{fig:phasediaferro} strongly
supports Gutzwiller's ideas on the importance of the local Hund's-rule
couplings.
At the same time the comparison of Figs.~\ref{fig:phasediaferro}a
and~\ref{fig:phasediaferro}b shows the importance of
band-structure effects which are the basis of the Stoner theory.
The ferromagnetic phase in the $U$-$J$ phase diagram is much bigger when
the density of states at the Fermi energy is large. Therefore, the
Stoner mechanism for ferromagnetism is well taken into account
by Gutzwiller's correlated-electron approach.

\begin{figure}[h]
\vspace*{-21mm}
  \begin{center}
    \epsfig{file=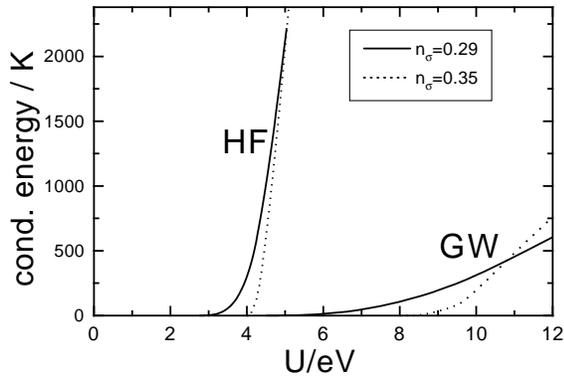,width=8cm}
  \end{center}
\vspace*{-50mm}
\caption{Condensation energy as a function of~$U$ for $J=0.2U$ for the
Hartree--Fock theory (HF) and the Gutzwiller wave function 
(GW) for $n/4=0.29$ (full lines) 
and~$n/4=0.35$ (dashed lines).}
\label{fig:condenenergy}
\end{figure}

In Fig.~\ref{fig:condenenergy}, we display the energy differences
between the paramagnetic and ferromagnetic ground states 
(``condensation energy'', $E_{\rm cond}$) as a function of
the interaction strength for $J=0.2U$. 
This quantity should be of the order of the Curie temperature which is in the
range of $100\,\text{K}-1000\,\text{K}$ in real materials. 
The Hartree-Fock--Stoner theory yields 
such small condensation energies only in the range 
of $U\approx 4\,$eV; for larger $U$, $E_{\rm cond}$ 
is of order $U$. In any case, the interaction parameter~$U$ 
has to be tuned very precisely to give condensation energies
which concur with experimental Curie temperatures~\cite{Slaterearly}.
In contrast, for the Gutzwiller-correlated wave function, we find
relatively small condensation energies $E_{\rm cond}=0.5\cdot 10^3\, {\rm K}$
even for interaction values
as large as twice the bandwidth ($U\approx 12\,$eV).
Moreover, the dependence of the condensation energy on~$U$ is rather weak
such that uncertainties in~$U$ do not drastically influence the
estimates for the Curie temperature.

These toy-model studies show that the Gutzwiller variational approach
contains all the fundamental elements for a
successful theory of itinerant ferromagnetism.
In the next section we show how to quantify this
statement for ferromagnetic nickel.

\section{Correlated band-structure of nickel}
\label{nickel}

\subsection{Discrepancies between experiment and SDFT}

Of all the iron group magnetic metals, nickel is the most celebrated case
of discrepancies between the results from experiment and from SDFT. From very
early on, the photo-emission data have indicated that the width of
the occupied part of the $d$~bands is 
approximately $W^*_{\rm occ} = 3.3\, {\rm eV}$~\cite{Eberhardt} 
whereas all SDFT 
results yield values of $W^*_{\text{occ, SDFT}}=4.5\, {\rm eV}$ or 
larger~\cite{Moruzzi,Eberhardt}.
Similarly, the low temperature specific heat data~\cite{Dixon} give
a much larger value of $N^*(E_{\rm F})$, 
the quasi-particle density of states at
the Fermi energy ($3.0$ versus $1.9$ states/(eV atom)), which indicates a
quasi-particle mass enhancement by a factor of approximately $1.6$.
Here, the Sommerfeld formula is used to convert the specific heat data;
the theoretical value follows directly from the quasi-particle 
band structure. Furthermore, very detailed 
photo-emission studies at symmetry points and
along symmetry lines of the Brillouin zone show discrepancies to SDFT
results for individual band-state energies which are of similar magnitude
as seen in the overall $d$~bandwidth. 

The studies also revealed even bigger
discrepancies in the exchange splittings of majority spin and minority
spin bands. The SDFT results give a rather isotropic exchange splitting of
about $600\, {\rm meV}$~\cite{Moruzzi,Eberhardt,Callaway}. In contrast, 
the photo-emission data show
small and highly anisotropic exchange splittings between $160\, {\rm meV}$ 
for pure $d(e_g)$~states such as $X_2$ and $330\, {\rm meV}$ for 
pure $d(t_{2g})$~states, the latter
value estimated from the exchange splitting of $\Lambda_3$ states along
$\Gamma$ to $L$~\cite{Donath,Guenthe}.
The much larger and much too isotropic exchange splitting of the SDFT results
has further consequences.
{\begin{enumerate}\parsep=0pt\itemsep=0pt
\item
The experimental magnetic moment of the strong ferromagnet Ni is
$\mu= 0.61 \mu_{\rm B}$; yet of relevance is its spin-only part
$\mu_{\text{spin-only}} = 0.55 \mu_{\rm B}$~\cite{Mook}. 
The SDFT result is $\mu_{\text{spin-only}} = 0.59 \mu_{\rm B}$~\cite{Moruzzi},
an overestimate related to the too large exchange splitting.
\item
the $X_2$~state of the minority spin bands lies below 
$E_{\rm F}$~\cite{Hopster}, whereas
all SDFT results predict it to lie above the Fermi 
level~\cite{Moruzzi,WangCallaway,Wohlfahrtreview}. As a consequence,
the SDFT Fermi surface exhibits {\sl two\/} hole ellipsoids 
around the $X$ point of the
Brillouin zone while in the de-Haas--van-Alphen experiments only {\sl one\/}
ellipsoid has been found~\cite{WangCallaway,Tsui}.
\item The strong $t_{2g}$-$e_g$ anisotropy is also reflected 
in the total $d$~hole spin density, i.e., in the observation 
that the $d$-hole part of the Ni magnetic
moment has 81\% $d(t_{2g})$ and 19\% $d(e_g)$ character~\cite{Mook}, 
while the SDFT results give a ratio of 74\% to 26\%~\cite{Jepsen}.
\end{enumerate}}
In the late 70's and early 80's various authors have investigated in how
far many-body effects improve the agreement between theory and 
experiment, see, e.g., Refs.~\cite{Cooke,Liebsch}.
For example, Cooke et al.~\cite{Cooke} introduced an anisotropic exchange
splitting as a fit parameter.

\subsection{Parameterization and minimization}

For the single-particle Hamiltonian~$\hat{H}_1$ in~(\ref{1})
we use a nine-orbital $s$, $p$, $d$ basis. The hopping matrix 
elements~$t_{i,j}^{\bbox{\sigma},\bbox{\sigma'}}$ are determined
from a least-square fit to energy bands obtained from
a density-functional-theory calculation for non-magnetic nickel.
The root-mean-square deviation of the $d$-band energies
is about $5\, {\rm meV}$. Details of this calculation
will be published elsewhere~\cite{BGWmega}.

The atomic Hamiltonian~$\hat{H}_{\rm at}$ in~(\ref{fullHat})
is restricted to only $d$~orbitals.
As a consequence, there are three independent interaction parameters
(we use the spherical-atom approximation), which can be expressed
either as Slater integrals $F^k$ ($k=0,2,4$) or as Racah 
parameters~$A$, $B$, $C$~\cite{Sugano}. 
We neglect any cubic crystal-field corrections to the interaction
parameters and, furthermore, put $C/B=4.5$, as suggested by the results
of ligand-field theory~\cite{Sugano}.
We find that $A\approx 10\, {\rm eV}$ and
$C\approx 0.1\, {\rm eV}$ are the choice which best reproduces
the experimental results; see below.

The number of multi-electron
states~$|\Gamma\rangle$ is $2^{10}-11$; because of the cubic site
symmetry, the number of independent (internal) variational 
parameters~$m_{\Gamma}$ reduces to approximately~200 for paramagnetic
and to approximately~400 for ferromagnetic solutions.
For the magnetic solutions, there exist a certain number of (external)
variational parameters~$\Delta_{\alpha}$ in the one-particle
wave function $|\Phi_0\rangle$. In particular, the~$\Delta_{\alpha}$
influence the spin-orbital densities $n_{i;b,\sigma}^0$ ($b=d,p,s$).
Among the $\Delta_{\alpha}$ 
are the ``exchange splittings'' of the individual orbitals $(b,\sigma)$
and $(b,-\sigma)$, i.e., these parameters determine the degree of magnetization
in $|\Phi_0\rangle$.
Most important are the exchange splittings for $d(e_g)$ and $d(t_{2g})$
orbitals, of minor importance are those for $s$ and $p$ orbitals.
Furthermore, the $d(e_g)$ vs.~$d(t_{2g})$ crystal-field splitting has to be
included in the set of $\Delta_{\alpha}$. Technically, each $|\Phi_0\rangle$,
which is a functional of the external variational parameters $\Delta_{\alpha}$,
defines a band structure and, consequently, a Fermi surface.
Therefore, calculating the particle densities $n_{i;b,\sigma}^0$
and the expectation values $\langle \Phi_0 | \hat{H}_{\rm eff}|\Phi_0\rangle$
implies momentum-space integrations up to the respective Fermi surface.
It should be noted that the introduction of the variational
parameters $\Delta_{\alpha}$ causes a charge flow between the $d$, and $s$
and $p$ states which has to be compensated by using appropriate ``chemical
potentials''. 

The optimum $|\Phi_0\rangle$ for a given set of $\{ \Delta_{\alpha} \}$,
$|\Phi_0^{\rm opt}\rangle_{\alpha}$ is the ground state of the effective
single-particle Hamiltonian~(\ref{Heff}) with renormalized
hopping matrix elements $\widetilde{t}_{i,j}$ which, through the $q$~factors,
are functions of the internal variational parameters $m_{\Gamma}$. 
Thus, $|\Phi_0^{\rm opt}\rangle_{\alpha}$ can be obtained from
a self-consistent procedure, starting with $q_{\bbox{\sigma}}=1$, i.e., with
$|\Phi_0^{\rm bare}\rangle_{\alpha}$. After minimization of
the internal variational parameters, the resulting values
for $q_{\bbox{\sigma}}$ define a new effective Hamiltonian from which a new
$|\Phi_0\rangle_{\alpha}$ is constructed. In this step, the Fermi surface
integrations have to be repeated. Self-consistency
is usually reached rather quickly, i.e.,
$|\Phi_0^{\rm opt}\rangle_{\alpha}$ is found after three to five iterations.

The global minimum, $|\Phi_0^{\rm opt}\rangle_{\rm global}$
is found by a search through the $N_{\alpha}$-dimensional
space of the external variational parameters~$\Delta_{\alpha}$.
This search can be sped up by first optimizing
with respect to the most important external variational
parameter which is the isotropic exchange splitting~$\Delta_d$.
In a second step, the anisotropy of the exchange splitting
is investigated, i.e., we introduce $\Delta_{e_g}$ and 
$\Delta_{t_{2g}}$, keeping the average $\Delta_d$ close to the value
of $\Delta_d^{\rm opt}$. The searches for $\Delta_d^{\rm opt}$
and for $\Delta_{e_g}^{\rm opt}$ and 
$\Delta_{t_{2g}}^{\rm opt}$ can be carried out starting
with $|\Phi_0^{\rm bare}\rangle$. Only then the self-consistency
procedure for $|\Phi_0^{\rm opt}\rangle$ has to be launched.

Typical energy gains are (in meV):
\begin{mathletters}
\begin{eqnarray}
E_0^{\rm bare} - E_0^{\rm bare}(\Delta_d^{\rm opt}) 
& \approx & 10-100 ,\\
E_0^{\rm bare}(\Delta_d^{\rm opt}) - 
E_0^{\rm bare}(\Delta_{e_g}^{\rm opt},\Delta_{t_{2g}}^{\rm opt})
& \approx & 5-10 , \\
E_0^{\rm bare}(\Delta_{e_g}^{\rm opt},\Delta_{t_{2g}}^{\rm opt})
- E_0^{\rm opt}(\Delta_{e_g}^{\rm opt},\Delta_{t_{2g}}^{\rm opt})
& \approx & 5-10 .
\end{eqnarray}\end{mathletters}%
The energy gains from the variations of 
$\Delta_{s}$ and $\Delta_{p}$ are of the order of $0.1\, {\rm meV}$.

At present, our description does not contain the spin-orbit coupling
which could be relevant in some transition metals.
This interaction contributes to the orbital moment $\mu_{\text{orbit}}
=0.06\mu_{\rm B}$ to the total magnetic moment.
The inclusion of the spin-orbit coupling poses no principle 
difficulties~\cite{BGWvoll} but it considerably enhances the numerical
complexity of the problem. 
Moreover, we may want to extend our basis set beyond the nine
bands $4s$, $4p$, and $3d$ to allow for a further relaxation of the
one-particle wave functions. 
Again, this is feasible but numerically much more costly.
Lastly, one may want to include some atomic $p$-$p$ and $p$-$d$ interactions.
All these extensions are left for future studies. 

\subsection{Comparison to experiments}

Typical results of our calculations 
for Ni are summarized in table~\ref{ourtable}.
They are obtained from the multi-band Gutzwiller method utilizing the
nine-orbital tight-binding model based on DFT energy band calculations
for non-magnetic Ni, and employing values of $A = 10\, {\rm eV}$
and $C = 0.1 \, {\rm eV}$, with $C/B = 4.5$.

The width of the $d$~bands is predominantly determined by $A$ (essentially
the Hubbard-$U$), via the values of the hopping reduction factors $q_{b,\sigma}$.
The exchange splittings and, consequently, the magnetic moment are strongly
influenced by~$C$ and only moderately by~$A$. The Racah parameter~$C$ 
causes the Hund's-rule
splitting of the $d^8$ multiplets; in the hole picture, $d^8$ is the only 
many-particle configuration which is significantly occupied (by $1.90$~electrons),
while $5.94$~electrons are in $d^9$, $0.89$~electrons are in $d^{10}$,
and $1.18$~electrons have $s$~or $p$~character.

Generally, the Gutzwiller results agree much better with experiment than
the SDFT results. This is the case for, (i), the $d$~bandwidth $W^*$, 
(ii), the value for $N^*(E_{\rm F})$, (iii), the positions of individual 
quasi-particle energies, (iv), the values of the exchange splittings,
(v), their $t_{2g}$-$e_g$~anisotropy, and, (vi),
the $t_{2g}/e_g$~ratio of the $d$~part of the magnetic moment. As a
consequence of the small $d(e_g)$~exchange splitting, the $X_{2\downarrow}$
state lies {\sl below\/} $E_{\rm F}$ and, thus, the Fermi surface 
exhibits only {\sl one\/} hole
ellipsoid around $X$, in nice agreement with experiment. 

\begin{figure}
\vspace*{120mm}
\smash{\vbox{\begin{table}
\begin{tabular}{l|clcc}
  & Exp. & Ref. & this work & SDFT  \\
\hline\\[-2ex]
$\mu_{\text{spin-only}}/\mu_{\rm B}$ & \hphantom{$+$}0.55 
& \protect\cite{Mook} & \hphantom{$+$}0.53 & \hphantom{$+$}0.59\\
$r(t_{2g}/e_{g})$ & \hphantom{$+$}4.3\hphantom{0} & \protect\cite{Mook} 
& \hphantom{$+$}5.1\hphantom{0} & \hphantom{$+$}2.8\hphantom{0} \\
$W^*_{\rm occ}= -\langle X_1\rangle/{\rm eV}$ & $-$3.3\hphantom{0} $\pm$ 0.2\hphantom{0} 
& \protect\cite{Eberhardt} & $-$3.14 & $-$4.86 \\
$W^*/{\rm eV}$ & 
\hphantom{$+$}3.5\hphantom{0} $\pm$ 0.3\hphantom{0} & 
\protect\cite{Eberhardt} & \hphantom{$+$}3.29 & \hphantom{$+$}5.20 \\
$N^*(E_{\rm F})/(\text{eV atom})$ & \hphantom{$+$}3.0\hphantom{0} 
& \protect\cite{Dixon} & \hphantom{$+$}2.6\hphantom{0} 
& \hphantom{$+$}1.9\hphantom{0} \\
$\langle X_3\rangle/{\rm eV}$ & $-$2.8\hphantom{0} $\pm$ 0.2\hphantom{0} 
& \protect\cite{Eberhardt} & $-$2.94 & $-$4.06 \\
$X_{2\uparrow}/{\rm eV}$ & $-$0.24 $\pm$ 0.02 & \protect\cite{Himpsel,Hopster}
& $-$0.22 & $-$0.56 \\
$X_{2\downarrow}/{\rm eV}$ & $-$0.06 $\pm$ 0.02 & \protect\cite{Himpsel,Hopster}
& $-$0.04 & $+$0.09 \\
$X_{5\uparrow}/{\rm eV}$ & $-$0.10 $\pm$ 0.02 & \protect\cite{Himpsel,Eastman}
& $-$0.14 & $-$0.22  \\
$X_{5\downarrow}/{\rm eV}$ & \hphantom{$+$}0.23 $\pm$ 0.04 
& \protect\cite{Himpsel,Eastman} & \hphantom{$+$}0.25 & \hphantom{$+$}0.34 \\
$\langle \Gamma_{12}\rangle/{\rm eV}$ & $-$0.4\hphantom{0} $\pm$ 0.1\hphantom{0} 
& \protect\cite{Eberhardt} & $-$0.64 & $-$1.02 \\
$\langle \Gamma_{25'}\rangle/{\rm eV}$ & $-$1.1\hphantom{0} $\pm$ 0.2\hphantom{0} 
& \protect\cite{Eberhardt} & $-$1.45 & $-$2.14 \\
$\langle L_{3}\rangle/{\rm eV}$ & $-$1.3\hphantom{0} $\pm$ 0.1\hphantom{0} 
& \protect\cite{Eberhardt} & $-$1.49 & $-$2.23 \\
$\Delta((1/2)\Gamma L)/{\rm eV}$ & \hphantom{$+$}0.21 $\pm$ 0.02 
& \protect\cite{Guenthe} & \hphantom{$+$}0.30 & \hphantom{$+$}0.60 \\
\end{tabular}
\caption{Comparison of experimental and theoretical results.
Parameters for the present work are:
$A=10\, {\rm eV}$, $C=0.1\, {\rm eV}$, and $C/B=4.5$.
The SDFT results are quoted from Ref.~\protect\cite{Moruzzi};
the result for $r(t_{2g}/e_{g})$ is taken from Ref.~\protect\cite{Jepsen}.
$\mu_{\text{spin-only}}$: spin contribution to the magnetic moment;
$r(t_{2g}/e_{g})$: ratio between $t_{2g}$ and $e_g$ contribution to the
magnetic moment $\mu_{\rm spin-only}$;
$W^*=X_{5\downarrow}-\langle X_{1}\rangle$: quasi-particle bandwidth;
$W^*_{\rm occ}$: occupied quasi-particle bandwidth;
$N^*(E_{\rm F})$: quasi-particle density of states at $E_{\rm F}$;
$\langle \ldots \rangle/{\rm eV}$: spin-averaged energies of the quasi-particle
states at various high-symmetry points of the Brillouin zone;
$X_{2,5\, \sigma}$: energies of the spin-$\sigma$ quasi-particle bands with
respective symmetries $X_2$/$X_5$ at the $X$ point;
$\Delta((1/2) \Gamma L)$: typical exchange splitting
of the highest-lying $\Lambda_3$ quasi-particle states halfway between
$\Gamma$ and $L$.}
\label{ourtable}
\end{table}}}
\vspace{-8mm}
\end{figure}

The large anisotropy of the exchange splittings is a result of our
ground-state energy optimization, which allows $\Delta_{t_{2g}}$ and 
$\Delta_{e_g}$ to be
independent variational parameters. We find
$\Delta_{t_{2g}} \approx 3\Delta_{e_g} \approx 800\, {\rm  meV}$. 
Note that these values enter
$|\Phi^{\rm bare}_0\rangle$ and are renormalized by factors 
$q_{b,\uparrow}\approx 0.7$, 
$q_{b,\downarrow}\approx 0.6$, 
when $|\Phi^{\rm opt}_0\rangle$ is reached.
This also implies that 
the width of the majority spin bands is about 10\% bigger than
that of the (higher lying) minority spin bands. It causes a further reduction
of the exchange splittings of states near $E_{\rm F}$, especially for those with
strong $t_{2g}$ character. 
Note that this band dispersion effect causes larger exchange
splittings near the bottom of the $d$~bands, e.g., $0.45\, {\rm eV}$ 
splitting of $X_1$ and $0.74\, {\rm  eV}$ splitting of $X_3$. 
There, however, the quasi-particle linewidths have
increased to $1.25\, {\rm eV}$ and $1.4\, {\rm  eV}$, respectively~\cite{Eberhardt},
so that an exchange splitting near the bottom of the $d$ bands could, so far, 
not be observed experimentally.
When we investigate the 
quasi-particle energies in the middle of the $d$ bands, 
such as $\Gamma_{12}$, $\Gamma_{25'}$, or $L_3$, we observe that our 
theoretical values lie consistently below the experimental ones. 
If we assume that these
states can be measured more reliably than those near the bottom of the
$d$~band, we may conclude that the value of $A$ could still be somewhat bigger 
than $10\, {\rm  eV}$. Studies with larger $A$~values are 
presently carried out.

We think that the large anisotropy originates from peculiarities
special to Ni with its almost completely filled $d$~bands and its fcc lattice
structure. Near the top of the $d$~bands, the $t_{2g}$~states dominate
because they exhibit the biggest hopping integrals to nearest neighbors,
$t^{(1)}_{dd\sigma}\approx 0.5\, {\rm eV}$. 
The $e_g$~states have $t^{(1)}_{dd\pi} \approx -0.3\, {\rm eV}$ to nearest
neighbors, and $t^{(2)}_{dd\sigma} \approx 0.1\, {\rm eV}$ to next-nearest
neighbors; the latter are small because of the large lattice distance
to second neighbors. The $e_g$~states also mix with the nearest-neighbor
$t_{2g}$~states with $t^{(1)}_{dd\pi}$-type coupling.
Therefore, the system can gain more band energy by avoiding occupation of
anti-bonding $t_{2g}$~states in the minority spin bands via large values of
$\Delta_{t_{2g}}$, at the expense of allowing occupation of 
less anti-bonding $e_g$ states via small $\Delta_{e_g}$~values. 
Note that this scenario is special to nickel and
should not apply to materials with a bcc lattice structure 
which have almost equal nearest and next-nearest neighbor
separations. 
Since the bands in nickel are almost completely filled,
the suppression of charge fluctuations actually reduces
the number of atomic configurations where the Hund's-rule coupling
is active. It is also in this respect that
nickel does not quite reflect the generic situation of
other transition metals with less completely filled $d$~bands; 
see Sect.~\ref{toymodel}.

The agreement between
theory and experiments on nickel shows that 
our multi-band Gutzwiller theory is a useful tool
for the theoretical description of the ground-state
properties of transition metals and their compounds.
In the next section we describe a way to calculate consistently
low-energy excitations in itinerant ferromagnets within our
variational approach.

\section{Spinwaves in itinerant ferromagnets}
\label{spinwave}

As in magnetic insulators, the elementary excitations in
itinerant ferromagnets are spinwaves.
Since ground-state properties are well described by
Gutzwiller-correlated wave functions it is highly desirable
to derive spinwave dispersions from the same variational
approach.

In fact, the variational principle can also be used to calculate
excited states~\cite{Messiah}. 
If $|\Phi\rangle$ is the ferromagnetic, exact
ground state with energy $E_0$, the trial states
\begin{equation}
|\Psi(q) \rangle = \hat{S}_q^- |\Phi\rangle
\end{equation}
are necessarily orthogonal to $|\Phi\rangle$, and provide
an exact upper bound to the first excited state with momentum~$q$
and energy $\epsilon(q)$
\begin{equation}
\epsilon(q) 
\leq E_{\rm s}(q) \equiv 
\frac{\langle \Psi(q) | \hat{H} | \Psi(q) \rangle}%
{\langle \Psi(q) | \Psi(q) \rangle} - E_0 \;.
\end{equation}
Here, $\hat{S}_q^-= (\hat{S}_q^+)^+= \sum_{l,b} \exp(-iql) 
\hat{c}_{l,b,\downarrow}^+\hat{c}_{l,b,\uparrow}^{\vphantom{+}}$
flips a spin from up to down in the system whereby it changes
the total momentum of the system by~$q$.
In this way, the famous Bijl-Feynman formula for 
the phonon-roton dispersion in superfluid Helium
was derived~\cite{Feynman}.
In the case of ferromagnetism the excitation energies~$E_{\rm s}(q)$ 
can be identified with
the spinwave dispersion if a well-defined spinwave exists at 
all~\cite{Buene2000}.
Experimentally this criterion is fulfilled for small momenta~$q$ 
and energies~$E_{\rm s}(q)$.

Unfortunately, we do not know the exact ground state or its energy
in general.
However, we may hope that the Gutzwiller wave function~$|\Psi_{\rm G}\rangle$
is a good approximation to the true ground state.
Then, the states
\begin{equation}
|\Psi_{\rm G}(q) \rangle = \hat{S}_q^- |\Psi_{\rm G}\rangle
\end{equation}
will provide a reliable estimate for $E_{\rm s}(q)$,
\begin{equation}
E_{\rm s}(q) \approx E_{\rm s}^{\rm var} (q)= 
\frac{
\langle \Psi_{\rm G} | \hat{S}_q^+ \hat{H} \hat{S}_q^-| \Psi_{\rm G} \rangle
}{
\langle \Psi_{\rm G} | \hat{S}_q^+ \hat{S}_q^-| \Psi_{\rm G} \rangle
}
-
\frac{
\langle \Psi_{\rm G} | \hat{H} | \Psi_{\rm G} \rangle
}{
\langle \Psi_{\rm G} | \Psi_{\rm G} \rangle
}
\;.
\end{equation}
Naturally, $E_{\rm s}^{\rm var} (q)$ does not obey any strict 
upper-bound principles.

The Hamiltonian~(\ref{1})
commutes with the operator for the total spin.
Thus, for $q=0$, $\hat{S}_{q=0}^-$ generates a state from the same
spin multiplet. Therefore, the spinwave at zero momentum
is the Goldstone mode of the ferromagnetic system, $E_{\rm s}(q=0)=0$.
Within the variational approach $E_{\rm s}^{\rm var}(q=0)=0$
is guaranteed if we choose $|\Psi_{\rm G}\rangle$ as an eigenstate
of the total spin operator. This is most easily accomplished
by a minor restriction of the variational space for the
atomic occupancies~\cite{Buene2000} with negligible influence
on the variational ground-state properties.

The actual calculation of the variational spinwave dispersion is
rather involved even in the limit of large dimensions.
We note, however, that explicit formulae are available~\cite{Buene2000}
which can directly be applied once the variational parameters have
been determined from the minimization of the variational ground-state energy.
As a first application, we present results for the two-band toy-model
of Sect.~\ref{toymodel}.

\begin{figure}
  \begin{center}
   \epsfig{file=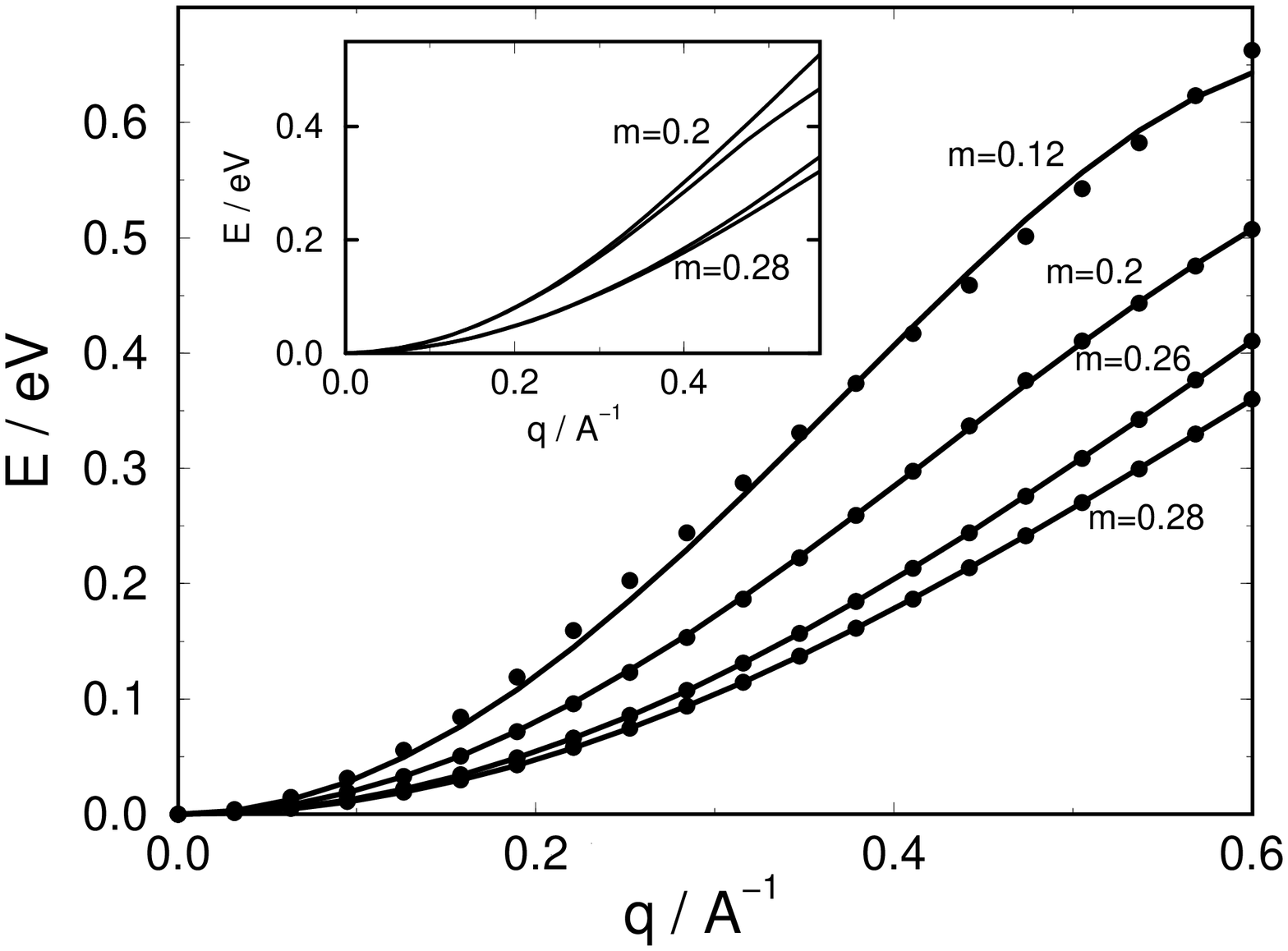,width=7.1cm}%
  \end{center}
\vspace{-4mm}
\caption{Variational spinwave dispersion in $x$~direction,
$E_{\rm s}^{\rm var}((q_x,0,0))$,
for the two-band model defined in Sect.~\protect\ref{toymodel};
$n/4=0.29$, $J=0.2U$, and the values 
$U/{\rm eV}=7.8,10,12,13.6$ correspond to $m=0.12,0.20,0.26,0.28$.
The lattice constant is $a=2.5\AA$. Inset: 
$E_{\rm s}^{\rm var}((q_x,0,0))$ and $E_{\rm s}^{\rm var}((q_x,q_x,0))$
for $m=0.2$ and $m=0.28$, respectively. The spinwave dispersion is 
almost isotropic.}
\label{fig:spinwaves}
\end{figure}

In Fig.~\ref{fig:spinwaves} we show the variational spinwave dispersion
in $x$~direction, $E_{\rm s}^{\rm var}((q_x,0,0))$,
for the model parameters $n_{\bbox{\sigma}}=0.29$, 
$J=0.2U$, and the four different
values $U/{\rm eV}=7.8,10,12,13.6$ which
correspond to a magnetization per band 
of $m=0.12,0.20,0.26,0.28$.
This quantity is defined as 
$0 \leq m=(n_{b,\uparrow}-n_{b,\downarrow})/2\leq n/4$.
Note that our last case corresponds to an almost complete 
ferromagnetic polarization.
The data fit very well the formula
\begin{equation}
E_{\rm s}^{\rm var}((q_x,0,0)) = Dq_x^2(1+\beta q_x^2) +{\cal O}(q_x^6) \; ,
\end{equation}
in qualitative agreement with experiments on nickel~\cite{nickelexp}.
The corresponding values $D=1.4\, {\rm eV}\AA^2$ and $D=1.2\, {\rm eV}\AA^2$ 
for $m=0.26$ and $m=0.28$, respectively, 
are of the right order of magnitude for nickel
where $D=0.43\, {\rm eV}\AA^2$. As lattice constant of
our simple-cubic lattice we chose $a=2.5\AA$.

As shown in the inset of Fig.~\ref{fig:spinwaves},
the dispersion relation is almost isotropic for
$q_x$~values up to half the Brillouin zone boundary~\cite{Buene2000},
in particular for large magnetizations. This is in contrast to
the strong dependence of the electron-transfer amplitudes~$t_{i,j}$
on the lattice direction. This implies for {\sl strong\/} ferromagnets
that the collective motion
of the local moments is similar to that of {\sl localized\/} spins
in an {\sl insulator}~\cite{Eschrig}. 
Such ferromagnetic insulators are conveniently
described by the Heisenberg model with exchange-interaction between
neighboring sites $\langle i,j\rangle$ on a cubic lattice,
\begin{equation}
\hat{H}_{\rm S} = - J \sum_{\langle i,j\rangle} \hat{\vec{S}}_i\hat{\vec{S}}_j
\; .
\end{equation}
For such a model one finds $D=2S J a^2$.
The length of the effective
local spins can be calculated from $|\Psi_{\rm G}\rangle$ 
as $S(S+1)\approx 0.95$
($S=0.6$) for $m\geq 0.20$~\cite{BGWvoll}. 
Therefore, $J\approx D/(1.2 a^2) $, which gives
the typical value $J= 0.17\, {\rm eV}$. For an estimate of the Curie
temperature~$T_{\rm C}$ we use the result from quantum Monte-Carlo
calculations~\cite{QMCspinmodels}
\begin{equation}
T_{\rm C} = 1.44 J S^2 
\end{equation}
for spins~$S$ on a simple-cubic lattice.
In this way we find $T_{\rm C}\approx 0.5 J=0.09\, {\rm eV}=
1\cdot 10^3 \, {\rm K}$.
This is the same order of magnitude as the condensation energy for these
values of the interaction, $E_{\rm cond} = 5 \cdot 10^2 \, K$,
see Sect.~\ref{toymodel}.
Given the arbitrariness in the relation between
$E_{\rm cond}$ and $T_{\rm C}$, and the application of the Heisenberg model
to our itinerant-electron system, we may certainly allow for 
difference of a factor two in these quantities.
Nevertheless, the results of this section clearly show that, (i),
$E_{\rm cond}$ gives the right order of magnitude for $T_{\rm C}$, and that,
(ii), the spinwave dispersion of {\sl strong itinerant\/} ferromagnets
resemble the physics of {\sl localized\/} spins.

\section{Conclusions}

Which scenario for itinerant ferromagnetism in transition
metals is the correct one?

For a long time this questions could not be answered conclusively.
Band theory along the lines of Slater and Stoner could be worked out
in much detail whereas a correlated-electron description
of narrow-band systems was lacking until recently.
In fact, van Vleck stated in the summary to his conference contribution
in 1953: ``The gist of this paper is that it would be highly desirable if good
methods of computing with~(c) (minimum polarity model)
could be developed.''
In 1963, Gutzwiller laid the foundations for a concise treatment of the 
problem,
and today we are able to draw definite conclusions
based on Gutzwiller-correlated multi-band wave functions.

Our results for nickel indicate that the van-Vleck--Gutzwiller scenario
is valid. Band theory does not appreciate the strong electronic
correlations present in the material which lead to the observed
renormalization of the effective mass, exchange splittings, bandwidths,
and Fermi surface topology. Moreover, as also covered by our approach,
charge fluctuations are small, and large local moments are present
both in the paramagnetic and the ferromagnetic phases.
Roughly we may say that the electrons' motion 
through the crystal leads to a ferromagnetic
coupling of pre-formed moments which eventually
order at low enough temperatures.
In this way, strong itinerant ferromagnets resemble ferromagnetic
insulators as far as their low-energy properties are
concerned: spinwaves exist which destroy the magnetic
long-range order at the Curie temperature. 

It is thus seen that today computational difficulties no longer obscure 
``the recognition in principle of the situation which confirms
closest to physical reality''(van Vleck, 1953~\cite{vanVleck}):
transition metals and their compounds 
are strongly correlated electron systems.
More than thirty-five years after their introduction
by Gutzwiller~\cite{GutzPRL63},
our studies clearly show that 
Gutzwiller-correlated multi-band wave functions
successfully describe the low-energy physics of these materials.

\section*{Acknowledgments}

We gratefully acknowledge helpful discussions with 
P.~van Dongen, H.~Eschrig, and D.~Vollhardt.
This project is supported in part by
the Deutsche Forschungsgemeinschaft under WE~1412/8-1.

\end{multicols}

\end{document}